\documentclass{article}

\usepackage{microtype}
\usepackage{graphicx}
\usepackage{subfigure}
\usepackage{booktabs} 

\usepackage{hyperref}

\usepackage[accepted]{latinx-icml-2021} 

\icmltitlerunning{Effects of personality traits in predicting grade retention of Brazilian students}

\begin{document}

\twocolumn[
\icmltitle{Effects of personality traits in predicting grade retention of Brazilian students}




\begin{icmlauthorlist}
\icmlauthor{Carmen Melo Toledo*}{usp}
\icmlauthor{Guilherme Mendes Bassedon*}{usp}
\icmlauthor{Jonathan Batista Ferreira*}{usp}
\icmlauthor{Lucka de Godoy Gianvechio*}{usp}
\icmlauthor{Carlos Guatimosim}{cs}
\icmlauthor{Felipe Maia Polo}{usp,ai}
\icmlauthor{Renato Vicente}{usp,exp}
\end{icmlauthorlist}

\icmlaffiliation{usp}{Universidade de São Paulo, São Paulo, Brazil}
\icmlaffiliation{ai}{Advanced Institute for Artificial Intelligence, São Paulo, Brazil}
\icmlaffiliation{cs}{ClearSale, São Paulo, Brazil}
\icmlaffiliation{exp}{Experian DataLab, São Paulo, Brazil}

\icmlcorrespondingauthor{Lucka de Godoy Gianvechio}{lucakggwp@gmail.com}
\icmlcorrespondingauthor{Jonathan Batista Ferreira}{jonathanbf@usp.br}
\icmlkeywords{Machine Learning, Personality traits, Education}

\vskip 0.3in
]



\printAffiliationsAndNotice{\icmlEqualContribution} 

\begin{abstract}



Student's grade retention is a key issue faced by many education systems, especially those in developing countries. In this paper, we seek to gauge the relevance of students' personality traits in predicting grade retention in Brazil. For that, we used data collected in 2012 and 2017, in the city of Sertãozinho, countryside of the state of São Paulo, Brazil. The surveys taken in Sertãozinho included several socioeconomic questions, standardized tests, and a personality test. Moreover, students were in grades 4, 5, and 6 in 2012. Our approach was based on training machine learning models on the surveys’ data to predict grade retention between 2012 and 2017 using information from 2012 or before, and then using some strategies to quantify personality traits' predictive power. We concluded that, besides proving to be fairly better than a random classifier when isolated, personality traits contribute to prediction even when using socioeconomic variables and standardized tests results.

\end{abstract}

\section{Introduction}



One of the most relevant problems of the Brazilian educational system is the grade retention rate, demanding critical attention from politicians and the academic community \cite{schwartzman2004challenges}. Since there are many students who end up repeating a school year or even dropping out \cite{nunes2014fatores}, some social measures are needed to prevent this behavior. In this work, we quantify the predictive power of personality traits in predicting grade retention. Since personality traits data are not traditionally used to that end or not even collected, we can help researchers and policy makers deciding how beneficial it would be if they start using that kind of features in their predictive models.





\section{Objectives and motivation}

The main objective of this paper is to assess, from a machine learning perspective, the predictive power of students' personality traits called "facets" \cite{soto2009ten}, measured in 2012, in predicting grade retention between 2012 and 2017 in the city of Sertãozinho (São Paulo, Brazil). In 2012, students were in grades\footnote{4º, 5º and 6º "anos".} 4, 5, and 6. In this work, we consider a student suffered grade retention if he/she advanced less than five grades between 2012 and 2017. That could happen when a student fails, temporarily drop out or due to any other possible reason that could interfere in such matter.

To achieve our objective, we adopted two strategies. At first, we test the performance of machine learning models trained only over those personality traits data, that have been collected in 2012, at evaluating the risk of future grade retention, trying to assess the hypothesis that the personality features are predictive by themselves. Additionally, we test the effects of combining personality traits features with more conventional data, that usually includes academic, demographic, socioeconomic features (such as standardized tests, school type, being public or private, ethnicity, mother education). We do that in order to test if there is a gain in the predictive power of the models with the addition of those new features (the personality traits). As for the evaluation method of the models, we adopted the area under the Receiver Operating Characteristic curve (ROC AUC). 


With that said, if we could verify our hypothesis that personality traits are indeed predictive for grade retention, it would be possible to implement more assertive tools, applied to the social context, making it possible to obtain more accurate measures of social vulnerability and having more precise interventions for avoiding grade retention among students, especially in developing countries. 

\section{Ethical concerns}

Since this work intersects with the social sciences, it is needed to reinforce some details. First, we must say that using machine learning algorithms to support/make decisions in a social context can be dangerous if the tools are misused. Any policy maker that intends to apply that kind of tool in real situations should be aware of possible biases and discriminatory behaviour of those algorithms. As for the practical side, it is recommended that any institution willing to use this paper's results follows the Five Pillars of Artificial Intelligence (AI) Ethics regarding the use of AI in schools and Education \cite{southgate2019artificial}.


\section{Related work}

Previous research that tried to find correlations between grade retention and students' characteristics in the Brazilian educational context are not new in the literature. \citet{nunes2014fatores}, for example, found correlations between school failure and low expectations of academic prospects, and correlations between “good perceptions about the school" and better academic perspectives. \citet{ortigao2013repetencia} observed that doing homework and having family support while going to school are important factors associated with decreased risk of grade retention. Furthermore, \citet{caluz2018papel} described the relation between personality traits and grade retention using econometric modelling and a dataset similar to ours. 

In our work, we add to the scientific literature by evaluating the statistical dependency of personality traits and grade retention from a machine learning perspective.

\section{Data}

The data used in this study was collected in the city Sertãozinho in the state of São Paulo - Brazil. The study was conducted by the “Laboratório de Estudos em Pesquisas e Economia Social" (LEPES/USP)\footnote{See \url{http://lepes.fearp.usp.br/}} and consists of field surveys with students in 2012 and the same students in 2017. In 2012, students were in grades\footnote{4º, 5º and 6º "anos".} 4, 5, and 6.


The information we used from the surveys can be split in two parts. The first part is a (i) personality test called Big Five Inventory, taken by students in 2012, which measured their scores in 10 “facets" presented in \citet{soto2009ten}, also described by \citet{piedmont1998revised}. The facets are: \textit{Activity}, \textit{Aesthetics}, \textit{Altruism}, \textit{Anxiety}, \textit{Assertiveness}, \textit{Compliance}, \textit{Depression}, \textit{Ideas}, \textit{Order}, and \textit{Self-Discipline}. The second part is composed by (ii) socioeconomic questions, that evaluate students' profiles in 2012 or in previous years, and mathematics and language standardized tests, taken by students in 2012. The tests were prepared using items of the National Institute for Educational Studies and Research Anísio Teixeira (INEP). A list with all the variables can be found in the supplementary material.

    
    
    
    
    
    
    
    
    
    

Our dataset in 2012 had roughly 4,900 students but only around 3,000 of them were interviewed in 2017. Moreover, considering that we only kept students with no missing information, our final analysis was made using 1888 subjects. Almost all students in 4$^{th}$ grade in 2012 were removed from the analysis, since most of them did not take the standardized tests due to bureaucratic reasons. 


\section{Methods}

\subsection{Exploratory data analysis}


For the exploratory step, we performed an analysis of pairwise mutual information of grade retention and personality traits with test scores features. This analysis helps us to understand the statistical dependence of grade retention and the features of interest. We use the Scikit-Learn\footnote{See \url{https://scikit-learn.org/stable/modules/generated/sklearn.feature_selection.mutual_info_classif.html} - accessed in 29/06/2021.} implementation of mutual information estimation, which relies on a non-parametric estimation method to measure the dependence of two random variables.

\subsection{Evaluation using predictive modelling and K-fold cross-validation}
For the rest of the sections, we used a logistic regression model and a XGBoost \cite{chen2016xgboost} classifier to measure the personality traits' predictive power. 


By performing a nested K-fold cross-validation procedure\footnote{Hyperparameter tuning details can be found in the supplementary material.} (outer $K=10$), we calculate the average AUC score using the $10$ iterations' test sets. We assess the models' performances in three cases: (i) using only the personality traits features (“Personalities only"), (ii) using all features other than personality traits (“Others only"), (iii) using the whole set of features. The first case will let us assess the predictive power of personality traits by their own, while the second and third cases will let us assess the predictive gain of adding personality traits to a set of more conventional features.

In order to draw error bars in our analysis, we employ the concept of standard error. Let $\textup{AUC}^{cv}$ be the vector containing AUC scores for each iteration of the outer K-fold cross-validation in the nested procedure. Thus, $\textup{AUC}^{cv}$ has length $K=10$. We use the following formula to estimate the standard error of our average AUC estimator:
\begin{equation}
    \widehat{SE}(\textup{AUC}^{cv}) =\frac{\widehat{SD}(\textup{AUC}^{cv})}{\sqrt{K}}
\end{equation} 

Where $\widehat{SE}$ is the estimated standard error, $\widehat{SD}$ is the $\textup{AUC}^{cv}$ sample standard deviation, and $K$ is the number of folds in the outer cross-validation loop ($K = 10$). 

\subsection{Gain analysis for including personality traits to the “Others only" features set}

To get more confidence and reduce error bars through variance reduction, we analyse the average AUC gain and the gain error bar, calculated in a similar way done for the ROC AUC scores. That is, instead of estimating average AUC scores and then taking the difference, we first take the difference of the individual $K=10$ AUC scores and then calculate their average. This procedure returns exactly the same difference but with lower standard error\footnote{This is an idea inspired by paired hypothesis tests, e.g., paired t-test.}. This is a valid approach since training and test sets are always the same in all combinations of features' sets and models in each of the $K=10$ iterations.
To identify whether this gain was significant we performed a paired one-tailed Student's t-test. In supplementary material, we tested pre-requisites of t-tests to verify if they are valid in our case.







\section{Results}

\subsection{Mutual information}

\begin{figure}[h!]
    \centering
    \includegraphics[width=0.48\textwidth]{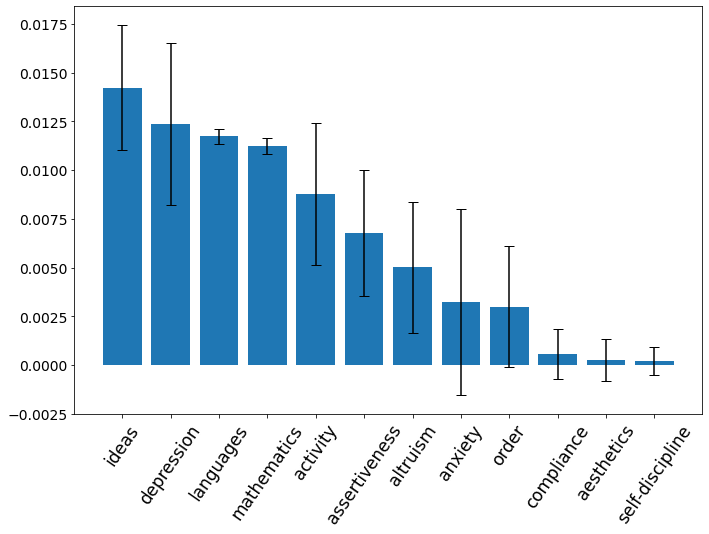}
    \caption{Average pairwise mutual information in Nats ($\pm$ standard deviation) of grade retention and one of the personality traits or test scores.}
    \label{fig:MutualInfo}
\end{figure}

We applied mutual information analysis to our data (Figure \ref{fig:MutualInfo}). Given that the used method depends on a random seed, we repeated this experiment 100 times and then recorded the average mutual information and standard deviation.

Depression and Ideas have shown the biggest dependencies with grade retention alongside with Language and Mathematics. Three personality traits (Aesthetics, Compliance and Self-discipline) showed little dependence with grade retention.

\subsection{Evaluation using predictive modelling and K-fold cross-validation}

\begin{figure}[h!]
    \centering
    \includegraphics[width=0.435\textwidth]{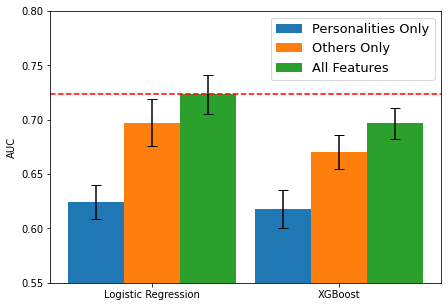}
    \caption{Bar plots of average cross-validation AUCs ($\pm$ standard error) with $K=10$. We combined three sets of features (Personalities only, Others Only and All features) and two models (logistic regression and XGBoost classifier). The red line displays the maximum average ROC AUC score achieved.}
    \label{fig:AUC bar plot}
\end{figure}

One can see in Figure \ref{fig:AUC bar plot} the bar plots of average cross-validation AUCs combining three sets of features (Personalities only, Others Only and All features) and two models (logistic regression and XGBoost classifier). It's notable that, in general, the logistic regression model performed better than XGBoost. 


In parallel, we see that for both types of models, the average AUC scores in the case of “Personalities only" reveal that, on average, the models performed fairly better than a random classifier, with scores that range from 0.60 to nearly 0.65. However, a clear result is that, in both logistic regression and XGBoost models, by adding personality traits to the “Others only" features set, we have a relatively good gain in the average AUC scores. In the following subsection, we will go into more details, with an analysis of this gain.

It is worth mentioning that, when looking at the error bars for both the “Others Only" and “All Features" features sets in Figure \ref{fig:AUC bar plot}, the gain is somewhat uncertain because the error bars intersect each other, leaving a margin for misinterpretation. For this reason we perform a analysis over the AUC gains.

\subsection{Gain analysis for including personality traits to the “Others only" features set}

The average AUC gains and their standard errors are represented in Table \ref{ttest}. Since the errors are lower than the average AUC gains, we can increase our confidence that there is an actual gain in performance with the inclusion of the personality traits to the “Others only" features set. 

The p-values in Table \ref{ttest} are obtained in a t-test. The hypotheses tested are: $H_0$ (that there is no average AUC gain by including personality traits to the “Others only" features set) and $H_1$ (that there is a positive gain). Since the p-value of the test in the logistic regression case is lower than 0.025, the null hypothesis $H_0$ can be rejected considering  5\% of significance. Considering XGBoost result, $H_0$ can be rejected considering a test of size 10\%. Thus, we have good reasons to think that the AUC gain by including personality traits to the “Others only" features set is positive.

\begin{table}[h]
 \centering
 \caption{Average AUC gain by including personality traits to the “Others only" features set. The p-values are calculated according to the hypotheses $H_0$ (that there is no average AUC gain by including personality traits to the “Others only" features set) and $H_1$ (that there is a positive gain). We have good evidence that the average AUC gain is positive.}
 \medskip
 \begin{tabular}{c|c|c|c}
 \toprule
 Model&Avg. AUC Gain & Std. error & p-value \\
 \midrule 
 Log. Reg. &$0.025$& $0.011$ & $0.023$ \\[.25em]
 XGBoost & $0.026$ & $0.015$ & $0.057$ \\
 \bottomrule
 \end{tabular}%
 \label{ttest}%
\end{table}%

Moreover, both logistic regression and XGBoost present average AUC gains around 0.025. That is an good increase relatively speaking, i.e., if we look at the logistic regression “Others Only" and “All Features" average scores, one can see it increased from approximately 0.7 to 0.725. Considering the minimum ROC AUC scores acceptable for machine learning models as being 0.5, we get that the increase was around $\frac{0.725-0.7}{0.7-0.5}=10\%$. The relative gain is even greater considering XGBoost.

\section{Conclusion}

In this work, we measure the predictive power of personality traits when predicting grade retention of Brazilian students. The use of personality traits to that end can assist more accurate public policies focused on decreasing grade retention rates, and it opens a window for further research to explore the gain of using those features as predictors. Our results show that we can use personality traits, besides socioeconomic indices and test scores, to have better machine learning models when predicting grade retention.

Future works might explore the policy of "progressão continuada"\footnote{Similar to automatic promotion. See \url{https://learningportal.iiep.unesco.org/en/glossary/automatic-promotion}}, which tries to lower evasion rates in Brazilian schools. It could be the case that some Sertãozinho's schools adopt that policy, and developing the analysis using this fact can lead to new discoveries. Also, breaking down the "Other only" features set can uncover insights in the future. 

Due to privacy reasons, we cannot share the dataset. The supplementary material and code can be found in \url{https://github.com/Lucka-Gianvechio/LatinX-Grade-Retention-Paper}.






\section{Acknowledgements}

We would like to thank the “Laboratório de Estudos em Pesquisas e Economia Social" (LEPES/USP) for sharing the dataset, and "Conselho Nacional de Desenvolvimento Científico e Tecnológico" (CNPq) for financially supporting Felipe during his master's degree.

\bibliography{bibliography}
\bibliographystyle{icml2021}

\end{document}